# DISCOUNT PUZZLE OF CLOSED-END MUTUAL FUNDS: A CASE OF BANGLADESH

**FARHANA RAHMAN**

Assistant Professor, Department of Organization Strategy & Leadership, University of Dhaka, Bangladesh

E-mail: farhanarahman@du.ac.bd

**Abstract**

The paper intends to perform a relevant study on the closed-end fund puzzle in the perspective of an emerging market. Quarterly data of 36 closed-end mutual funds traded in Dhaka Stock Exchange are collected over the sample period of 2016 to 2019. Dependent and independent variables are mapped down by exploring previous researches. Weight of top 10 investments, fund size, fund age, fund maturity, turnover and dividend yield are taken as explanatory variable to analyze the impact on CEF discount. A fixed effects panel regression is performed on the data set with few diagnostic tests to ensure the reliability of the analysis conducted. The results show that, the variable fund size and fund maturity have a significant positive and turnover has a significant negative impact on CEF discount while the impact of weight of top 10 investments, dividend yield and fund age are found insignificant.

**Keywords:** Closed-End Funds (CEF), Discount Puzzle, Net Asset Value (NAV), Bangladesh (BD).

## 1. INTRODUCTION

Trading of closed-end funds (CEF) at a discount from asset's net asset value (NAV) has been a long-talked issue in the literature of financial economics. The objective of the paper is to perform a relevant study on discount puzzle in Dhaka stock exchange (DSE). Similar kind of studies have been conducted in several developed markets though, but never been performed in the perspective of an emerging market like Bangladesh.

Closed end funds (CEF) are investment vehicle that accumulate funds from investors, invest the funds in portfolio of securities and manage the funds to generate return for investors. CEFs accumulate the capital through initial public offering of shares, which are in fixed numbers and are traded in the stock exchanges. The investment vehicles are a perfect example that represent observable market value and fundamental value of securities, thus has been largely studied to research the phenomena of efficient market hypothesis (EMH) which stipulates that market value and intrinsic value of assets should be equal. Pratt (1966) first took the initiative of using CEFs to compare the market price and value of underlying assets and found that closed-end funds are traded at price below its net asset value, a phenomenon widely known as the discount puzzle.

Theoretically, the price of assets with similar risk-return distribution cannot be traded at different prices. If traded, price discrepancy will cause arbitrage opportunity that is the asset can be purchased at price lower than its original value. Over the years, several researches have been done all over the world trying to explain the behavior of deviation between CEF units and their NAVs. Some researchers have talked about rational framework where factors like illiquidity, tax or management fees affect the deviation while others have put forward behavioral aspects like investor sentiment to find an explanation of the age-old puzzle. However, it has not been possible to provide a single explanation to the puzzle. Studies done so far could only explain a portion of the puzzle. Since the very inception of mutual fund industry in Bangladesh (BD), the CEFs has been trading at a deep discount from NAV. The paper aims to study the puzzle in an emerging market perspective.

## 2. OBJECTIVES

The objective is to identify few factors that determine the discount of CEF and measure their effects in the perspective of an emerging market. The following answers could be found after completion of the analysis:

i. What factors influence the discount at which Bangladeshi CEFs continue to be traded?

ii. Are the results consistent with finance theories and previous research findings?

## 3. MUTUAL FUND INDUSTRY OF BANGLADESH

The study is conducted in the time frame of 2016 to 2019. During that time, there were 37 closed end funds trading at Dhaka stock exchange managed under 9 asset management companies (AMC). The funds were trading continuously at a discount from NAV over time. Though the event is very common in financial markets worldwide, explanations for existence of the puzzle in the perspective of BD has been untouched. One of the reasons of no study so far can probably be the fairly newness of the industry. Though the industry started its journey in 1980, it received its momentum at some point in 2010. Institutional investors account for more than 65% (DSE Shareholding Status Report, 2019) of total Asset under management (AuM) which means the industry failed to attract retail investors. The retail investors in Bangladesh are still heavily dependent on depositing money in banks instead of putting money in mutual funds despite the fact that the contribution of the banking sector in economic development can be disrupted because of the existence of significant amount of non-performing loan (NPL) in the banking system (Hamid & Rahman, 2020).

Closed-end funds in India trade at approximately 5-15% discount to NAV on stock exchanges (Mahesh, 2019). The situation is quite different Bangladesh. The closed-end funds in Bangladesh traded on an average of 30% discount from NAV over the years of 2016 to 2019. CEFs in BD has been continuously trading at deep discount from NAV since the very inception of the industry. Figure 1 represents average discount from NAV of CEFs trading in Dhaka Stock Exchange. All the 37 funds are included from the first quarter of 2016 to the end of the quarter of 2019.

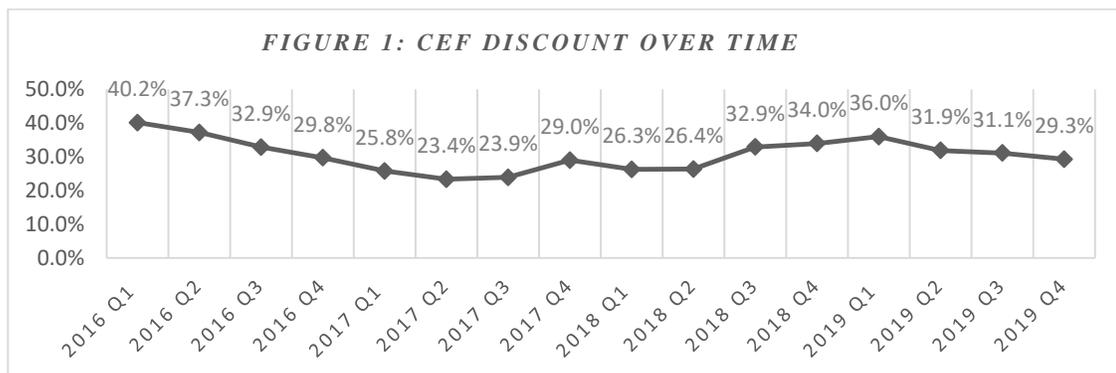

*Figure 1: CEF Discount Over Time*

## 4. LITERATURE REVIEW

A closed-end mutual fund is a collective investment undertaking where fixed amount of capital is raised through initial public offering to finance its investments. Unlike open end funds, closed-end funds have fixed capitalization meaning that CEFs have fixed number of shares to be traded on stock exchanges or OTC markets. The fund achieves diversification through investing in different classes of financial assets rather than a single instrument. The share price of closed-end funds reflects their market values rather than net asset values. As demand and supply force determine the price of closed-end funds, share prices are typically higher or lower than the NAV constituting the premiums or discounts of CEFs. Empirically it has been seen that CEF shares sell at prices that is in discrepancy with the market prices of its underlying assets. This phenomenon is widely known as the discount puzzle of mutual funds. There might be some funds that sale at premium, but the general norm shows a discount of 10% to 20% (Rosenfeldt & Tuttle, 1973).

The enigma of deviation has been studied by different academic communities. Two different approaches have come into consideration to solve the decade old puzzle. Efficient Market Hypothesis (EMH) is the first approach. Asset prices should reflect its fundamental value (Fama, 1970). However, the existence of various market imperfections in the real world like management fees, taxes, asset portfolio characteristics can be the contributing factor behind the divergence of CEF share price and its NAV. The second approach is Investor Sentiment Hypothesis (ISH). According to ISH, the discount or premium arises due to the emotional differences of investors which in turn effects the share price and asset valuation (Pratt, 1966; Boudreaux, 1973; Lee, Shleifer and Thaler, 1990).

Lee, Shleifer and Thaler (1991) in their analysis about discount puzzle portrayed four periods describing the lifecycle of closed-end fund puzzle.

1. At the launch of new funds, CEF's are usually traded at premium from their intrinsic value. During the period of floatation, this overvaluation is of almost 10%. According to Lee, Shleifer and Thaler (1991) the reason behind this appreciation of CEF share price is startup costs and underwriting costs. These costs are removed from the proceeds reducing the fund NAV. Basically investors are interested to pay premiums for new CEFs.

2. Within 120 days of initial trading, the premium vanishes, and the shares start to trade at an average of 10% discount. This observation is also confirmed by the studies conducted by Weiss, Lehn and Malmquist (1989) where they conducted the analysis on 64 closed end funds and showed that there has been a gradual loss in the cumulative index-adjusted return after 120 trading days.

3. There are variations in discounts depending on time and funds. But the discounts are mean reverting. Lee et al. *(*1991) also showed high correlation of discounts between funds. Studies conducted by Bodurtha, Kim and Lee (1995) also showed similar results.

4. In the 4th stage of life cycle of discount puzzle, the authors showed that discounts tend to vanish and share prices converge to NAV during the period of termination, conversion to open-end funds or liquidations.

The presence of discount puzzle raises questions to the advocates of market efficiency and traditional finance paradigm regarding the fact that why discount or premium occurs and what are the factors contributing to them. The rational framework depicted illiquidity effects, tax effects, management fee and management performance as components to explain the discount puzzle (Malkiel, 1977). However, researchers also explored alternative dimensions within the behavioral finance literature where investor sentiment is a prominent factor contributing to the puzzle.

In the rational framework, management fee has been put forward as a component explaining discount. The idea behind the hypothesis is that, when a manager having no or reduced stock picking and timing skills charges fees, the stocks of the fund will be traded at less than its NAV. Ross (2002) argued in favor of the argument. However, as the management fee for closed-end funds in Bangladesh is fixed, the variable cannot be used to explain any cross-sectional variation in fund discount.

Management performance is another facet to consider. Past performance has been studied by Malkiel (1977) who found no evidence of relationship between discount and past performance. Roenfeldt and Tuttle (1973) found a weak relationship when they studied about contemporaneous performance. However, according to the research results of Lee, Shleifer, and Thaler (1990) discounts are found to be weakly related to future NAV performance.

Evidence of relationship between diversification and discount has been found by Hjelström (2007). The 2008 study results of Chan, Wan Kot, and Li measured diversification by Herfindahl index using the weight of top 10 stocks and found a negative association between Herfindahl index and discount of closed-end funds traded in Chinese market referring to the fact that discounts decline with lower level of diversification.

According to Pontiff (1996) CEFs that pay large dividends tend to have smaller discount. The high dividend paying funds are mispriced to a lesser degree. Dividends paid by a CEF or company has signaling effects but there are opposing views depending on either it is interpreted positively or negatively by investors. According to some researchers, high dividends represent positive outlook on future cash flows, thus favored by investors (Starks, Yoon, 1995). Malkiel (1977) argued that investors benefit from high dividend payout ratios. The study results found in Swedish market showed that, the higher the dividends paid, the higher the discount (Egerot & Hagman, 2011) which contradicts the expected results.

Study of ownership structure of closed-end funds has added another dimension to the literature. When the ownership composition is concentrated highly, it leads to higher discount according to Barclay, Holderness, and Pontiff (1993). Similar kind of results are found in Chinese market (Chan, Wan Kot & Lee, 2008). The logic behind the argument is that the majority owners may have other preference over minority investors and create agency problems. As a result, the equity claim is punished by investors through higher discounts.

Several researches have associated the fund's age with discounts. It is found that though funds trade at premium after their initial listing, they tend to fall to discount within 120 days of initial trading and the discount vanishes over time as the fund reaches close to its termination period (Lee, Shleifer & Thaler, 1991). The notion behind argument is that very few funds could beat the market even though they are launched with the motive to beat. If the funds cannot beat the market, they trade at discount.

Tax effects are portrayed to be an important factor to solve discount puzzle according to rational framework. Investors are required to pay taxes on CEF's realized capital gains, thus funds which have large capital gains usually trade at discount to NAV (Malkiel, 1977). Tax timing explanations are proposed by Brickley, Manaster and Shallheim (1991) and Kim (1994) to clarify discount puzzle. On the other hand, Chen, Rui, and Xu (2004), in their research paper, did not find capital gain taxes as an important factor in explaining discount puzzle. In USA, dividends are highly taxed resulting in CEFs trading in discounts from NAV.

Assets liquidity can be a major component causing discount or premium. If assets are infrequently traded, it causes barriers to arbitrage triggering discount (Gemmill and Thomas, 2002). Chan, Wan Kot and Lee (2008) in their studies showed that discount of CEF is positively related to the size of open-end funds issued by the same fund house.

A possible solution of discount puzzle might be found through exploring investor sentiment which covers up the possible shortcomings of thoughts of neoclassical school. It is the irrational investors, known as noise traders in the literature, whose sentiment and actions push the CEF price away from closed-end fund NAV.

## 5. METHODOLOGY

The paper used previous works in different markets on different time frames to create a new and time demanding applicable work on Bangladeshi market. Potential variables contributing to CEF discounts are mapped down by studying previous research works. Quarterly data of Bangladeshi CEF population is studied during the period 2016 to 2019. ***Standard Linear Panel Regression Analysis*** is conducted to assess the effect of the explanatory variables on CEF Discount.

In this paper, we are dealing with short panel. Data of 36 CEFs over 16 quarters is analyzed. Six independent variables are chosen to assess the impacts of those variables on the explained variable CEF Discount.

$$DISCOUNT_{it} = \alpha + \beta_2 TOP\_10\_STOCK_{it} + \beta_3 DIV\_YIELD_{it} + \beta_4 FUND\_SIZE_{it} + \beta_5 FUND\_AGE_{it} + \beta_6 FUND\_MATURITY_{it} + \beta_7 TURNOVER_{it}$$

The assumption of all the funds not being same calls for the use of ***Fixed effects model*** for the analysis. By allowing to have own intercept value, this model allows individuality. The term fixed effect is because although the intercept may differ across funds, the intercept does not vary over time making any individual fund attribute time invariant (Gujarati, 2004). It means that all the closed end funds are having own intercept value.

***Random effects model*** is also known as error component model. The difference of random effects model from fixed effects model is that it assumes a common intercept $\alpha$ for cross-sectional units which does not vary over time or cross-sectional units. That is all the 36 funds can have common mean value for the intercept. The random variable $\varepsilon_i$ is steady over time but varies across entities (Brooks, 2014). ***Hausman test*** is applied to identify which model (Fixed effects or random effects) is appropriate.

Then heteroscedasticity test is performed. Due to the presence of heteroscedasticity in the model, panel regression is again estimated with clustered robust standard error. ***Sargan-Hansen Test*** is later applied to determine between fixed effects and random effects because Hausman test only takes default standard errors in estimation while providing decision between fixed effects or random effects model. The test of muticollinearity and normality has been conducted.

## 6. DATA DESTRIPTION

### 6.1 Market and Time Frame

In the year 2020, there were 37 listed Closed-end Mutual funds in the stock exchange of Bangladesh. 36 funds out of overall population of 37 funds are taken into account for analysis purpose. The sample time frame for our analysis is from 2016 to 2019. Even though, information is accessible for few mutual funds before 2016, many CEFs were missing necessary data for the analysis. That is why we restrict our examination to the sample time frame. Quarterly data from the year 2016 to 2019 is collected to perform an up to date analysis on the issue.

### 6.2 Variable Description

In this study, impact of several independent variables on the CEF discount is to be assessed. The fund discount is computed in the following way by using quarterly data on individual fund prices in the market and their respective NAVs for the sample time frame.

$$Dis_{i,t} = \frac{NAV_{i,t} - Share\ Price_{i,t}}{NAV_{i,t}} \qquad (Ross, 2002)$$

Here $Dis_{it}$ refers to the fund discount of fund $i$ at the end of quarter $t$. Positive value of the variable indicates discount. On the other hand, negative value of the variable refers to premium. Share $Price_{it}$ in the above equation indicates market price of fund share $i$ at the end of quarter $t$, while $NAV_{it}$ indicates net asset value per share at the end of quarter $t$ of fund $i$.

The following variables are considered to have a potential impact on fund discount, thus taken for analysis purpose in the paper. The variables are chosen by studying the previous research works conducted on the same research field in different markets. Most of the variables are chosen from the research paper published by Chan, Wan Kot and Lee (2008) which studies portfolio concentration and closed-end fund discounts in the Chinese market.

**6.2.1 Weight of top 10 stocks (TOP10_STOCK)** — defined as the weight of top 10 stocks in the fund. It measures the level of concentration of the stocks in the portfolio. The variable is estimated as the weight of top 10 stocks among the listed securities divided by the total market value of investment in the listed securities. It is to be mentioned that most of the CEFs in Bangladesh do not provide information about the non-listed investments in their portfolios. The higher the value of the variable, the lower the level of diversification in the fund portfolio. In this case, if the better stock picking skill generates the benefit that is greater than the cost of having concentrated portfolio, a higher TOP10_STOCK will be followed by a reduced discount. On the contrary, we expect to have a positive relationship between the variable and fund discount if the investors of mutual funds do not prefer stock concentration.

**6.2.2 Dividend yield (DIV_YIELD)**: Defined as paid dividend divided by market price (Kim & Song, 2010). The variable is expected to have a positive relationship with discount. The higher the dividend paid by the CEF, the better off the investors. As a result, investors reward high dividend paying fund with higher demand, thus lower discount.

**6.2.3 Fund size (FUND_SIZE):** Computed as the natural logarithm of net asset value times units outstanding of the fund (Kim & Song, 2010). Fund size may portray two opposite effects on discount. A larger fund is followed by higher liquidity and consequently have lower trading cost. It means there persists a negative relationship between Fund Size and CEF discount. On a second thought, larger funds can make managers less inclined to stock selection causing the investors to push down the CEF price, thus creating positive association between the variables.

**6.2.4 Fund age (FUND_AGE):** This variable is estimated by taking the natural logarithm of number of days since the listing of the fund in the stock exchange. It is expected that funds with a longer listing history become more well-known to the investors in the market and have a smaller discount.

**6.2.5 Fund maturity (FUND_MATURITY):** The variable is measured by the natural logarithm of the number of days left for fund redemption. It is expected that, the variable will show a positive relationship with CEF discount as when a fund goes close to its maturity date or liquidation or conversion date, the discount vanishes.

**6.2.6 Turnover (TURNOVER):** this ratio is determined as the number of units traded over a quarter divided by the number of CEF units outstanding (Kim & Song, 2010). Higher turnover implies greater liquidity, which is why we foresee a negative association between the variable Turnover and discount from the funds. This ratio is used as a proxy for the liquidity of the fund.

## 7. RESULTS

### 7.1 Descriptive Statistics

Table 1 portrays the descriptive statistics of the dependent and independent variables chosen for the study. The data set consists of 16 quarters over the sample period of 2016 to 2019 across 36 CEFs trading in the stock exchange. The data set is unbalanced panel. Natural logarithm of fund size, fund age and fund maturity are taken to normalize the data for analysis purpose (Rahman, 2017). From the table it can be seen that CEFs traded at an average discount of 30%. It can be seen that there is wide fluctuation of discount from -85% to 63%. The mean fund age is 2226 days or 6.2 years. The average fund size is approximately 1580 million BDT. Considerably higher degree of dispersion is noticed in the variable fund size and fund age.

*Table 1: Descriptive Statistics*

| Variable | Mean | Standard Deviation | Min | Max |
|---|---|---|---|---|
| Discount | .3044425 | .1820007 | -.8506132 | .6367113 |
| TOP_10_STOCK | .6214146 | .2204097 | .2695618 | 2.254995 |
| DIV_YIELD | .0176762 | .0104723 | 0 | .0477941 |
| FUND SIZE (million BDT) | 1580 | 1410 | 179 | 8780 |
| FUND_AGE (days) | 2226.377 | 946.1789 | 22 | 4013 |
| FUND_MATURITY (days) | 3827.913 | 1346.637 | 61 | 6271 |
| TURNOVER | .173169 | .3250432 | .0000261 | 2.763141 |

## 7.2 Test of Multicollinearity

A correlogram has been generated by taking all the independent variables to detect the presence of strong association among the independent variables. The variables are not correlated to the extent of causing the problem of multicollinearity in the regression model.

*Table 2: Test results of Multicollinearity*

|  | DIV YIELD | FUND SIZE | FUND AGE | FUND MATURITY | TOP_10_STOCK | TURNOVER |
|---|---|---|---|---|---|---|
| DIV YIELD | 1.00 |  |  |  |  |  |
| FUND SIZE | -0.39 | 1.00 |  |  |  |  |
| FUND AGE | 0.19 | 0.09 | 1.00 |  |  |  |
| FUND MATURITY | -0.22 | 0.48 | 0.00 | 1.00 |  |  |
| TOP 10_STOCK | 0.06 | 0.22 | -0.23 | -0.03 | 1.00 |  |
| TURNOVER | -0.04 | -0.32 | 0.06 | -0.43 | -0.11 | 1.00 |

## 7.3 Hausman test

Both fixed effects and random effects panel regression are conducted. To choose between fixed effects and random effects, Hausman test is conducted for both models. Table 3 shows the test results. The p value is 0.00 which is highly significant. Thus, we can reject the null hypothesis that random effects model is appropriate.

*Table 3: Hausman Test Results*

| Hausman (1978) specification test | **Chi-Sq. Statistic** | **Prob.** |
|---|---|---|
|  | 48.443 | 0.0000 |

## 7.4 Test of Heteroscedasticity

After specifying that panel regression with fixed effects is appropriate for the analysis, Heteroscedasticity test is conducted on fixed effects regression model. The results of Modified Wald Test for Groupwise Heteroskedasticity shows that the probability value for Chi-Sq statistics is significant. It refers to the rejection of the null hypothesis which is residuals are homoscedastic. It indicates that the problem of heteroskedasticity is present in the model.

*Table 4: Test Results for heteroscedasticity*

| Modified Wald Test for Groupwise Heteroskedasticity | **Chi-Sq. Statistic** | **Prob.** |
|---|---|---|
|  | 1038.55 | 0.0000 |

## 7.5 Presentation of Regression Results (with Clustered Robust SE)

To account for the problem of heteroscedasticity, fixed effects model is estimated with clustered robust standard errors. Before reporting the results of fixed effects panel regression model, it is important to mention that, in the previous section Hausman test was conducted to choose between fixed effects or random effects model. Hausman test takes the default standard errors in consideration while giving decision about model appropriateness, not the robust standard errors. To see if inclusion of robust standard errors has altered the decision regarding appropriateness of fixed effects model, a test of overidentifying restrictions: fixed vs random effects is conducted. The null hypothesis for the test is same as Hausman specification. The test results are summarized in Table 5. Sargan- Hansen statistic shows that the probability value of Chi-Sq. Statistics is 0.00 indicating the rejection of random effects model. So, Hausman test and Sargan-Hansen test give the same results that fixed effects model is appropriate for the analysis purpose.

*Table 5: Test of overidentifying restrictions: fixed vs random effects*

| *Sargan-Hansen statistic* | **Chi-Sq. Statistic** | **Prob.** |
|---|---|---|
| | 66.683 | 0.0000 |

The regression results for fixed effects with robust standard error is summarized below in table 6. The variables Dividend yield, Fund size, Fund Maturity and Fund turnover were found significant at different level of significance. Weight of top 10 investments found to have insignificant impact on discount of CEFs trading in DSE.

*Table 6: Fixed Effects Model Results with Clustered Robust SE*

| Discount | Coef. | Robust St. Err. | p-value |
|---|---|---|---|
| TOP10_STOCK | .02 | .039 | .609 |
| DIV_YIELD | 4.082 | 1.463 | .008*** |
| FUND_SIZE | -.139 | .051 | .01** |
| FUND_AGE | .035 | .035 | .328 |
| FUND_MATURITY | .14 | .06 | .025** |
| TURNOVER | -.088 | .03 | .005*** |
| Constant | 1.753 | 1.361 | .206 |
| | | | |
| Within R-squared | 0.280 | F-test | 12.374 |
| Number of obs | 493.000 | Prob > F | 0.000 |
| *** p<.01, ** p<.05, * p<.1 | | | |

The result of fixed effects regression with times effects is summarized in table 7. Cross-sectional variations that is Fund level variations arising from unobserved factors are already captured by using fixed effects in panel regression. However, to measure time series variations, times effects are added. It checks if part of the variations in discount can be explained by time trends.

*Table 7: Fixed Effects Model with Time Effects and Clustered Robust SE*

| Discount | Coef. | Robust St. Err. | p-value |
|---|---|---|---|
| TOP10_STOCK | -.021 | .02 | .303 |
| DIV_YIELD | 1.17 | 1.379 | .402 |
| FUND_SIZE | .239 | .074 | .003*** |
| FUND_AGE | .036 | .033 | .287 |
| FUND_MATURITY | .203 | .055 | .001*** |
| TURNOVER | -.053 | .031 | .095* |
| Constant | -6.531 | 1.478 | 0*** |
| | | | |
| Within R-squared | 0.473 | F-test | 51.803 |

| Number of obs | 493.000 | Prob > F | 0.000 |
|---|---|---|---|
| *** p<.01, ** p<.05, * p<.1 | | | |

When the time effects are taken into consideration, the within R squared which was approximately 28% with the fixed effects model without period effects increased to approximately 47.3%. Apart from that, the Dividend Yield variable, which was significant before, became insignificant implying there the impact of dividend yield on fund discount is statistically 0. There is another difference to be noticed which is the coefficient of fund size was negative in the fixed effects model. But when time effects are added, the coefficient sign of the variable flips and turn to positive.

## 8. ANALYSIS

All the variables included in the study were taken from previous researches that were relevant in explaining the phenomena. Efforts have been put forward to see if the variables remain relevant while put in the context of Bangladeshi market. Explanations for the variables are provided considering pattern of mutual fund industry in Bangladesh and previous research findings.

### 8.1 Fund size

The fund size is found significant in fixed effects panel estimate. The variable shows negative relationship with fund discount (Table 6). But when time effects are taken into consideration, the coefficient sign flips (Table 7). Thus, results from fixed effects model with time variation seems to be appropriate for explanation because adding time effects remove effects of any sort of cyclical or market wide factors on all funds. Chan, Wan Kot and Lee (2008) on their study of Chinese market hypothesized that there will be higher discounts for large funds as it is likely that fund managers will invest in lots of stock without proper stock selection and fund will be less diversified in true sense, and found the relationship significant.

The stock market in Bangladesh is very volatile. As the market cannot be predicted properly, in times of market slumps, it is difficult to get out of the stock positions already held. The problem becomes more prominent for bigger funds. Apart from that as the funds get bigger in size, it becomes difficult to manage the funds efficiently. The fund managers may engage into less stock selection causing agency cost between CEF investors and fund managers. As a result, the investors punish by pushing the price down of the closed end fund units, thus increasing the discount.

### 8.2 Fund Maturity

As a CEF reaches to periods of maturity, its share price converges to NAV, discount vanishes and fulfills no arbitrage paradigm. The above theory is used to hypothesize the fact that the longer the fund's maturity or redemption date, the higher the discount of the CEF at which it trades. As the fund reaches close to its maturity date, discount reduces.

The variable fund maturity is found significant. It is also positively associated to discount. The finding is consistent with the findings of Lee, Shleifer and Thaler (1991) and Brickley and Schallheim (1985). Lee, Shleifer and Thaler (1991) described four important pieces of discount puzzle.

### 8.3 Turnover

Turnover measure is taken as a proxy for fund liquidity. From the fixed effect regression estimates including and excluding time effects, both ways the variable is found to be significant and negatively related with fund discount. The regression coefficient for the variable is .05 (Table 6 & 7). It means that, as the turnover increases by 1%, CEF discount decreases by approximately 5%. High stock liquidity is expected to be associated with low discount. The results are consistent with previous literature. In their study of Chinese market Chan, Wan Kot and Lee (2008) found higher demand for discount for a less liquid market.

## 9. CONCLUSION

In this paper, efforts have been put forward to identify and analyze the impacts of different explanatory variables on discount of closed-end funds trading in Dhaka Stock Exchange (DSE). 36 out of the whole population of 37 closed-end mutual funds are analyzed at quarterly intervals over a period of 2016 to 2019. Different econometric tools and methods are used for analyzing the panel data.

In each model, within the set of possible explanatory variables, fund size, fund maturity and turnover are found to have significant impact on CEF discount. Should investor sentiments be taken into account, the model would have more explanatory power. In other words, the market is filled with noise traders. Rational framework alone is not sufficient to explain the discount puzzle in market setting like Bangladesh where the market is volatile and unpredictable, and most investors are irrational and lacks financial literacy.

Fund Maturity is found in line with previous research having a significant positive impact on CEF discount in Bangladesh. This is in line with life cycle of CEF puzzle. As the fund starts trading in the market, it is traded at a premium. Afterwards the funds move to discount as the expectation is that it is not possible for fund managers to constantly beat the perfect market portfolio. As the fund moves closer to its redemption period, the discount eventually vanishes and CEF price converges to its NAV. This explanation is consistent with the fact that the longer the time-period towards fund redemption date, the higher the discount.

Fund Size is also found to have statistically significant impact on the dependent variable fund discount. The variable portrayed a positive association. The effect of Fund Size on CEF Discount can be explained both positively and negatively according to previous research. In Bangladeshi market positive association seemed to be dominant. It indicates that investors perceive large funds as inefficient as it might give rise to agency costs between investors and fund managers due to the fund managers lesser engagement in stock selection.

The variable Turnover is found significant and negatively related to CEF discount representing that stronger liquidity is associated with lower the discount. It is in line with expectations and research findings. Other variables like measures of diversification, fund age and dividend yield, are found insignificant in the market setting of Bangladesh but variables were impactful in other market settings.

Apart from the rational components, some behavioral aspects should be taken into account while trying to understand the discount puzzle. There is lack of confidence and financial literacy among investors in overall market. General investors have a tendency to follow herd instincts rather than rational decision-making process. There is suboptimal asset class and product diversification. Also, the regulatory and legal framework is not efficient. Due to all these issues, investors have developed a not so positive sentiment about the overall market, let alone the MF industry. Bangladesh is a bank dependent economy despite the upsurge of nonperforming loans which has contributed to increased concern over the industry (Rahman & Hamid, 2019). The Mutual Fund Industry has not been able to attract large investor base despite 40 years of operation.

All the above factors mentioned above together are directly or indirectly causing the deviation of CEF prices from its underlying asset values. For the mutual fund industry to grow, the confidence of the investors must be restored. And the confidence will be restored by increasing investor awareness; and ensuring transparency and governance in the market by enforcing strict regulatory and legal action.

The paper may not be able to solve the age-old discount puzzle, however it tried to detect some of the pieces of the puzzle. Apart from the identification of significant variables like fund size, maturity and turnover that impacts the CEF discount; several problems, shortcomings and discrepancies in the Bangladeshi market and some features of investor's behavior and sentiment are put forward that may be deemed as a contributing factor to the discount puzzle. It is expected that the findings of the paper will encourage more rigorous research on the study area by academicians and practitioners and provide better insights to asset management companies about factors impacting the discount of CEF.